\documentclass[aps,11pt,amsmath,amssymb,amsfonts,notitlepage]{revtex4-1}
\usepackage{graphicx}
\usepackage{enumerate}

\begin{document}
\title{Local spacetime effects on gyroscope systems}

\author{Mattias N.\,R. Wohlfarth}
\email{mattias.wohlfarth@desy.de}
\affiliation{II. Institut f\"ur Theoretische Physik und Zentrum f\"ur Mathematische Physik, Universit\"at Hamburg, Luruper Chaussee 149, 22761 Hamburg, Germany}
\author{Christian Pfeifer}
\email{christian.pfeifer@desy.de}
\affiliation{II. Institut f\"ur Theoretische Physik und Zentrum f\"ur Mathematische Physik, Universit\"at Hamburg, Luruper Chaussee 149, 22761 Hamburg, Germany}

\begin{abstract}
We give a precise theoretical description of initially aligned sets of orthogonal gyroscopes which are transported along different paths from some initial point to the same final point in spacetime. These gyroscope systems can be used to synchronize separated observers' spatial frames by free fall along timelike geodesics. We find that initially aligned  gyroscope systems, or spatial frames, lose their synchronization due to the curvature of spacetime and their relative motion. On the basis of our results we propose a simple  experiment which enables observers to determine locally whether their spacetime is described by a rotating Kerr or a non-rotating Schwarzschild metric.
\end{abstract}

\maketitle

\section{Introduction}

Gravity in general relativity is described by the curvature of spacetime. So a question that naturally arises is how observers may determine this curvature in their local neighborhood~\cite{AudLae} by its effects on physical objects, like point particles and spinning gyroscopes. The most commonly discussed such effect is the relative acceleration between freely falling point particles near an observer's position. Mathematically, this is described by the Jacobi equation and hence sourced by the so-called electric components of the Riemann tensor~\cite{Nichols:2011pu}.

In this article we will describe the effects of curvature on relatively moving gyroscope systems in detail. Concretely, we will consider two sets of gyroscopes that define an orthogonal system of spatial axes; these are synchronized at some initial point of spacetime; then they are transported along different paths to a nearby final point where their axes are compared. It will turn out that the so-called magnetic components of the Riemann tensor cause a loss of synchronization of the gyroscope systems which adds to non-gravitational effects coming from their relative velocity and relative acceleration. Mathematically speaking, this generalizes the statement of the path-dependence of parallel transport on curved spacetimes into a statement about the path-dependence of Fermi--Walker transport. Physically speaking, we will show that the loss of synchronization of initially aligned gyroscope axes can be interpreted as the relative rotation of observers' spatial frames induced by gravity and their relative motion.

Experiments and calculations that study the effects of spacetime curvature on gyroscopes have been considered before, e.g. in \cite{AudLae,Nichols:2011pu,Will:2002ma,Adler:1999yt, Schiff}. In contrast to earlier theoretical approaches, however, we construct a fully realizable synchronization and comparison procedure for the gyroscopes through timelike free-fall propagation. Moreover, our calculation is fully covariant and applies to arbitrary spacetimes; neither does it require a time-space split nor the metric perturbation theory of the parametrized post-Newton formalism. 

As an application of our new results we will discuss a local experiment which enables observers to decide whether the spacetime they are living in is of Kerr or Schwarzschild type, i.e., whether it possesses angular momentum or not. This is achieved locally without referring to observers or other quantities defined at infinity or using spacetime perturbation theory. We simply need to study the curvature induced relative rotation of gyroscope systems for local observers.

We will begin in section~\ref{sec:vectrans} by reviewing some mathematical concepts needed to analyze the curvature effects on gyroscope motion. In addition, we will develop the Fermi--Walker transport of gyroscopes along non-differentiable worldlines which is required to generalize the path-dependence theorem. In section~\ref{sec:theorems} we will describe how observers may use gyroscope systems to synchronize their spatial frames. Our main result Theorem~1 (which we will prove in full mathematical detail in the appendix \ref{sec:proof}) then connects the desynchronization of gyroscope systems  to the magnetic components of the Riemann tensor. We will then derive Theorem~2  that interprets our results in terms of relative local rotations of observers' spatial frames of reference. In section \ref{sec:kerr} we will apply our findings to demonstrate that a simple local gyroscope experiment can distinguish between Kerr and Schwarzschild spacetime.  We will conclude with a discussion in section \ref{sec:disc}. 

\section{Transport of vectors and gyroscopes in spacetime}\label{sec:vectrans}
Before we study how curvature effects gyroscopes, we review and develop some mathematical preliminaries. First, in section~\ref{sec:tp}, we recall the notion of covariant Taylor series on metric manifolds $(M,g)$ and the standard theorem about the connection between parallel transport along different paths and the Riemann curvature tensor. Then, in section~\ref{sec:basicgm}, we clarify the basic properties of a gyroscope and review its motion through spacetime. Finally, in section~\ref{sec:gyrondw}, we derive new results on the motion of gyroscopes along non-differentiable worldlines. The latter are needed for a precise description of the motion of the gyroscope systems considered in section \ref{sec:theorems}.

\subsection{Covariant Taylor expansions and path-dependence}\label{sec:tp}
The covariant Taylor expansion on metric manifolds $(M,g)$ is a way to expand tensors around a point  $p$ in $M$ such that the coefficients of the Taylor series are covariant objects; details can be found in~\cite{covT}. Here, we only discuss the first order expansion of vector fields since this is all we will require below. 

Consider a curve $\gamma: t\mapsto\gamma(t)$ through $\gamma(0)=p$ and a vector field $X(\gamma(t))$ along this curve. Let $\{p_\mu(\gamma(t))\}$ with $\mu=0, \dots , 3$ be a basis of $T_{\gamma(t)}M$ which is parallelly transported along $\gamma$, i.e., $\nabla_{\dot\gamma}p_\mu = 0$. Observe that this implies 
\begin{equation}\label{eq:sma}
\nabla_{\dot\gamma} X = \dot \gamma(X^\mu)\ p_\mu\,.
\end{equation}
Now we express $X(\gamma(t))$ in the parallelly transported basis and then Taylor expand the components to equate
\begin{equation}
X(\gamma(t))=X^\mu(\gamma(t))p_\mu(\gamma(t))=\Big[X^\mu(\gamma(0))+t\ \dot\gamma (X^\mu)(\gamma(0))\Big] p_\mu(\gamma(t))+\mathcal{O}(t^2)\,.
\end{equation}
Combining these formulae shows that this expansion is in fact a covariant Taylor expansion with 
\begin{equation}\label{eq:covt}
X^\mu(\gamma(t))= X^\mu(\gamma(0))+t\ (\nabla_{\dot\gamma} X)^\mu(\gamma(0))+\mathcal{O}(t^2)\,.
\end{equation}
This type of expansion will be used repeatedly in the appendix \ref{sec:proof} where we will prove Theorem~1 of section~\ref{sec:theorems}. Another small but important fact following  from~(\ref{eq:sma}) is that the components of a parallelly transported vector field with respect to a parallelly transported basis are constant along the path.

A well-known mathematical theorem states that the Riemann curvature of metric manifolds $(M,g)$ measures the difference of the parallel transports of an initial vector along two different paths to the same final point. 

\begin{figure}[h]
\includegraphics[width=2.7in]{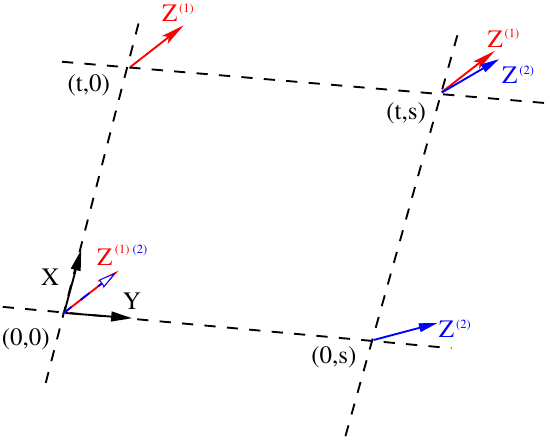}
\caption{\label{fig:curv}Parallel transport of a vector along two different paths.}
\end{figure}

More precisely, consider two commuting vector fields $X$ and $Y$ on $M$ with $[X,Y]=0$. Now moving from $p$ first along an integral curve of $X$ for parameter distance $t$, then along an integral curve of $Y$ for parameter distance $s$ (path 1) reaches the same final point as moving from~$p$ first distance $s$ along $Y$, then distance $t$ along $X$ (path 2). The final points for different $t,s$ form a surface ${\gamma: (t,s)\mapsto\gamma(t,s)}$ with $p=\gamma(0,0)$ and partial derivatives $\dot\gamma(t,s)=X(\gamma(t,s))$ and $\gamma'(t,s)=Y(\gamma(t,s))$, see figure \ref{fig:curv}. Let $Z\in T_pM$ be an initial vector at the point $p\in M$. We write $Z^{(1)}$ and $Z^{(2)}$ for the vector fields generated from this vector by parallel transport along the respective paths 1 and 2. With this notation the theorem states~\cite{Wald}:
\begin{equation}\label{eq:pd}
Z^{(2)}(\gamma(t,s))-Z^{(1)}(\gamma(t,s)) = ts\, (R(X,Y)Z)(p) + \mathcal{O}((t,s)^3)\,.
\end{equation}

The central result of this paper, Theorem~1, will be a generalization of this theorem. It shows that the path-dependence of the Fermi--Walker transported spin axes of gyroscope systems is related to the Riemann curvature tensor and their relative motion. We will show that this has a nice physical interpretation: gravity induces a relative rotation on all spatially separated frames of reference and thus on all physical objects.

\subsection{Basic gyroscope motion}\label{sec:basicgm}
The physics of a gyroscope is characterized by an intrinsic angular momentum vector field~$S$ defined along the timelike worldline $\gamma:t\mapsto \gamma(t)$ with unit normalized four-velocity, $g(\dot\gamma,\dot\gamma)=-1$. The field~$S$ satisfies two requirements:
\begin{enumerate}[\;(1)]
\item $S$ is spatial in the gyroscope's reference frame, i.e., $g(S,\dot \gamma)=0$; 
\item no torque is applied to the gyroscope; following \cite{Weinberg} this means $\nabla_{\dot \gamma}S \sim \dot\gamma$.
\end{enumerate}
By differentiating the first condition along the worldline and projecting the second condition on the direction $\dot\gamma$, one finds the transport equation for $S$:
\begin{equation}\label{eq:gyro}
\nabla_{\dot\gamma}S = g(\nabla_{\dot\gamma}\dot\gamma,S)\, \dot\gamma\,.
\end{equation}
We emphasize that the unit normalization of $\dot\gamma$ is crucial for this equation to hold; it is not invariant under rescalings of $\dot\gamma$. The above transport law for the intrinsic angular momentum of a gyroscope  is known as Fermi--Walker transport.

The gyroscope transport equation implies that the normalization $g(S,S)>0$ stays constant along the worldline $\gamma$. This fact can be used to show that the axis of rotation defined by the normalized vector field $e_S = S/\sqrt{g(S,S)}$ satisfies the same equation as $S$. Now consider a system of three gyroscopes with initially orthonormal rotation axes $e_\alpha$ for $\alpha=1, \dots,  3$. Then the initial orthonormality $g(e_\alpha,e_\beta)=\delta_{\alpha\beta}$ of this system is preserved along the worldline, since
\begin{equation}
\dot\gamma (g(e_\alpha,e_\beta)) = 2 \,g(\nabla_{\dot\gamma}e_{(\alpha},e_{\beta)}) = 2 \, g(\nabla_{\dot\gamma}\dot\gamma,e_{(\alpha}) \, g(e_{\beta)},\dot\gamma) = 0\,.
\end{equation}
Hence an orthonormal system of gyroscopes always will stay orthonormal, no matter how it moves. In particular the system could be arbitrarily accelerated and rotated.

The last ingredient missing before we can describe the synchronization of spatial frames is the transport of gyroscopes along non-differentiable worldlines. The difficulty here is to guarantee that the gyroscope axes stay spatial with respect to their actual worldline tangent.

\subsection{Gyroscopes on non-differentiable worldlines}\label{sec:gyrondw}
An observer with a gyroscope could decide at some point $t_0$ in time to move it around in his laboratory. In an idealized situation, the worldline $\gamma$ of this gyroscope that followed the worldline of the observer for $t<t_0$ would then non-differentiably branch off at $\gamma(t_0)$. Hence
\begin{equation}
e_0 = \lim_{t \nearrow t_0}\dot\gamma(t) 
\neq \lim_{t\searrow t_0}\dot\gamma(t) = f_0\,.
\end{equation}
Here the gyroscope transport equation (\ref{eq:gyro}) can only be applied on the domains $t<t_0$ and $t>t_0$. There it ensures that the intrinsic angular momentum $S$ always stays spatial, i.e., orthogonal to~$\dot\gamma$, which is realized by infinitesimal pure Lorentz boosts in the planes spanned by~$\dot\gamma$ and the acceleration $\nabla_{\dot\gamma}\dot\gamma$. 

However, at $t_0$ the notion of space abruptly changes. There the final intrinsic angular momentum $S^<(t_0)$ reached along $t<t_0$ which is orthogonal to $e_0$ has to be mapped by hand into a corresponding value $S^>(t_0)$ orthogonal to $f_0$ that may serve as initial condition for further transport for $t>t_0$. The relevant map is the unique finite pure Lorentz boost $\Lambda$ that maps $e_0\stackrel{\Lambda}{\curvearrowright}f_0$ without involving any spatial rotation. 

In order to find this map we consider the generators for boosts in the plane spanned by $\langle e_0, f_0\rangle$, where we write $f_0=f_0^\mu e_\mu$ with respect to an orthonormal basis $\{e_\mu\}$ with $\mu=0\dots 3$. These generators have components
\begin{equation}
\omega^\mu{}_\nu = 2\,e_0^{[\mu} f_0^{\rho]}\,\eta_{\rho\nu} = 2\,\delta_0^{[\mu} f_0^{\rho]}\,\eta_{\rho\nu}\,.
\end{equation}
Exponentiation $\Lambda=\exp(\omega \lambda)$ with parameter $\lambda$ determined so that $\Lambda(e_0)=f_0$ then yields the Lorentz transformation matrix
\begin{equation}
\Lambda^\mu{}_\nu = \left[\begin{array}{cc}f_0^0&f_{0\,\beta}\\f_0^\alpha&\delta^\alpha_\beta+\frac{f_0^0-1}{|\vec f_0|^2}f_0^\alpha f_{0\,\beta} \end{array}\right]
\end{equation} 
with respect to the basis $\{e_\mu\}$, where $|\vec f_0|^2=\delta_{\alpha\beta}f_0^\alpha f_0^\beta$ and spatial indices $\alpha$ are lowered with $\delta_{\alpha\beta}$. 

The application of this Lorentz transformation to an intrinsic angular momentum vector $S^<=S^{<\,\alpha} e_\alpha$ orthogonal to $e_0$ yields after some rewriting
\begin{equation}\label{eq:boost}
S^> = S^< +\frac{g(f_0,S^<)}{1-g(f_0,e_0)}(f_0+e_0)\,.
\end{equation}
This formula shows that the difference $S^>-S^<$ indeed lies in the plane $\langle e_0, f_0\rangle$. Moreover, it is not difficult to check that $S^>$ is now spatial with respect to the new time direction, $g(f_0,S^>)=0$, and has the same length as before the Lorentz boost, $g(S^>,S^>)=g(S^<,S^<)$.
 
\section{Relative gyroscope rotations}\label{sec:theorems}
This is the central section of this article, where we will analyze in detail the spacetime effects on gyroscope systems by considering the path-dependence of gyroscope, or Fermi--Walker, transport. We will show that two systems of initially synchronized gyroscopes that form aligned orthonormal spatial frames of reference gradually lose their synchronization. The effects come from non-vanishing magnetic components of the Riemann tensor and the relative velocity and acceleration of the gyroscope systems. Our results are presented in the form of two theorems. Theorem~1 in section~\ref{sec:ex} establishes the details of gyroscope desynchronization, and Theorem~2 in section~\ref{sec:localrot} provides a physical interpretation in terms of measurable relative rotations of the gyroscope axes and spatial observer frames.

\subsection{Curvature effects on gyroscope transport}\label{sec:ex}
Consider a freely falling observer who prepares two aligned copies of a system of three gyroscopes with axes that represent her spatial orthonormal frame. Let a second observer move along an arbitrary timelike worldline in some small distance to the first. In order to equip the second observer with a synchronized gyroscope system, the first  initiates the free fall of one of the two aligned gyroscope systems to the second. Then both observers propagate with their gyroscope systems through spacetime. After some time they wish to compare their gyroscope systems. To do so, the first observer initiates another free fall of her remaining gyroscope system to the second observer, who now compares the axes of both gyroscope systems. In general, it will turn out that the gyroscopes have lost their synchronization. We will see that to first non-trivial order this desynchronization is sourced not only by the Riemann curvature tensor of spacetime but also by contributions depending on the relative velocity and relative acceleration between the two observers.

Mathematically, we describe the spacetime effects on gyroscope systems by a two-surface $(t,s)\mapsto \gamma(t,s)$ with parameters $t$ and $s$. For fixed $t$ or $s$ this surface can be understood as a family of curves with respective tangents $\dot\gamma=\partial_t \gamma$ or $\gamma'=\partial_s\gamma$. We have $[\gamma',\dot\gamma]=0$.
The freely falling first observer moves on the geodesic $t\mapsto \gamma(t,0)$ with $(\nabla_{\dot\gamma} \dot\gamma)_{|(t,0)}=0$; this can be parametrized so that $g(\dot \gamma,\dot \gamma)_{|(t,0)}=-1$. The second observer moves along the timelike trajectory $t\mapsto \gamma(t,s)$ for some fixed $s$ without further special properties. The free fall of the gyroscope systems from the observer on $\gamma(t,0)$ to the observer on $\gamma(t,s)$ takes place along timelike curves $s\mapsto \gamma(t,s)$ for fixed $t$ which are again geodesics with $\nabla_{\gamma'}\gamma'=0$.  

\begin{figure}[h]
\includegraphics[width=2.7in]{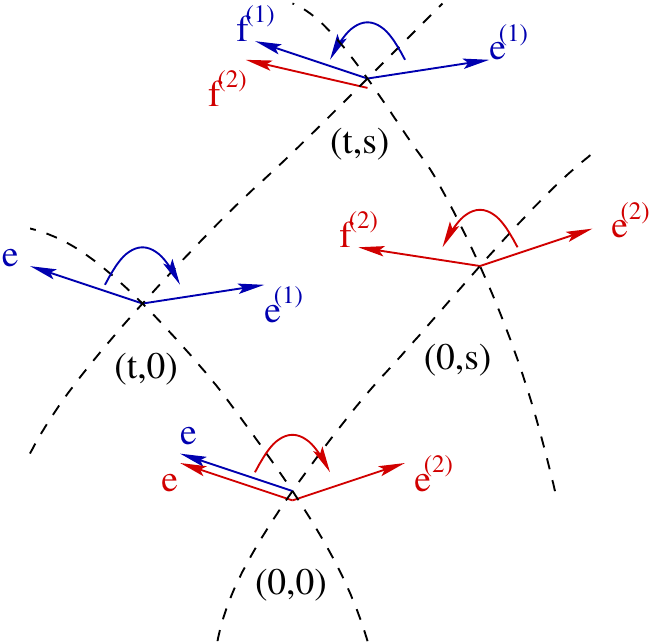}
\caption{\label{fig:ex}Motion of initial gyroscope axes $e$ at $\gamma(0,0)$ to final gyroscope axes $f^{(1)}$ and $f^{(2)}$ at $\gamma(t,s)$ along different paths: ${}^{(1)}$ via $\gamma(t,0)$, and ${}^{(2)}$ via $\gamma(0,s)$. Finite Lorentz boosts changing orthogonality to $\dot\gamma$ into orthogonality to $\gamma'$ are indicated by $\curvearrowright$, the reverse change by $\curvearrowleft$.}
\end{figure}

The gyroscope transport can now be performed as illustrated in figure \ref{fig:ex}. The freely falling observer with worldline $t\mapsto \gamma(t,0)$ prepares two identical spatial orthonormal gyroscope systems with axes $e_\alpha(0,0)$ at $\gamma(0,0)$. One copy is transported freely falling along $s\mapsto \gamma(0,s)$ to a second observer with worldline $t\mapsto \gamma(t,s)$; since $e_\alpha(0,0)$ is orthogonal to $\dot\gamma(0,0)$, this motion must involve a finite Lorentz boost $e_\alpha(0,0)\mapsto e_\alpha^{(2)}(0,0)$ to keep the orthogonality between the gyroscope axes to their worldline tangent $\gamma'(0,0)$. Now the transport equation (\ref{eq:gyro}) can be applied. When the second observer receives the gyroscope at $\gamma(0,s)$ another Lorentz boost has to be applied to ensure that the gyroscope axes become orthogonal to the second observers worldline $e_\alpha^{(2)}(0,s)\mapsto f_\alpha^{(2)}(0,s)\perp \dot\gamma(0,s)$. This procedure defines the initial synchronization of  the two observers' gyroscope systems $e_\alpha(0,0)$ and $f_\alpha^{(2)}(0,s)$. Now both observers transport their gyroscope systems for some time $t$. The resulting $f_\alpha^{(2)}(t,s)$ provide the final axes orientations of the second observer that must be compared to those of the gyroscope system $e_\alpha(t,0)$ of the first  observer at $\gamma(t,0)$. To do this the latter must be moved to the point $\gamma(t,s)$; this involves the finite Lorentz boost $e_\alpha(t,0)\mapsto e_\alpha^{(1)}(t,0)\perp \gamma'(t,0)$, followed by a free fall to $e_\alpha^{(1)}(t,s)$ and another finite Lorentz boost $e_\alpha^{(1)}(t,s)\mapsto f_\alpha^{(1)}(t,s)\perp \dot\gamma(t,s)$.
 
The two paths along which the copies of gyroscope systems are transported can be summarized as follows,
where $\overset{X}{\longrightarrow}$ indicates the use of the gyroscope transport equation along the tangent vector~$X$ and $\overset{\Lambda}{\curvearrowright}$ the application of a finite Lorentz boost $\Lambda$:
\begin{eqnarray}\label{eq:paths}
{}^{(1)}: &\;& e_\alpha(0,0) \overset{\dot\gamma(t,0)}{\longrightarrow} e_\alpha(t,0) \overset{\Lambda_{11}(t)}{\curvearrowright} e_\alpha^{(1)}(t,0) \overset{\gamma'(t,s)}{\longrightarrow} e_\alpha^{(1)}(t,s)\overset{\Lambda_{12}(t,s)}{\curvearrowright} f_\alpha^{(1)}(t,s) \\ 
{}^{(2)}: &\;& e_\alpha(0,0) \overset{\Lambda_{21}}{\curvearrowright} e_\alpha^{(2)}(0,0) \overset{\gamma'(0,s)}{\longrightarrow} e_\alpha^{(2)}(0,s) \overset{\Lambda_{22}(0,s)}{\curvearrowright} f_\alpha^{(2)}(0,s) \overset{\dot\gamma(t,s)}{\longrightarrow} f_\alpha^{(2)}(t,s) \nonumber\,.
\end{eqnarray}
Note that the resulting gyroscope systems $f_\alpha^{(1)}(t,s)$ and $f_\alpha^{(2)}(t,s)$ of the two observers that are to be compared now are both orthogonal to $\dot\gamma(t,s)$. The result of this transport of frames can be stated as follows:

\vspace{6pt}\noindent\textbf{Theorem 1.} \textit{Let $\gamma:(t,s)\mapsto\gamma(t,s)$ be a two-dimensional surface embedded in a spacetime $(M,g)$ so that $(\nabla_{\dot\gamma}\dot\gamma)_{|(t,0)}=0$ with $g(\dot\gamma,\dot\gamma)_{|(t,0)}=-1$, and $\nabla_{\gamma'}\gamma'{}_{|(t,s)}=0$. Let $e_\alpha(0,0)\perp \dot\gamma(0,0)$ be an orthonormal spatial frame which is Fermi--Walker transported (as a gyroscope system) into frames $f_\alpha^{(1)}(t,s)$ or $f_\alpha^{(2)}(t,s)$ at $\gamma(t,s)$ along either path~(1) or path~(2) stated above in equation (\ref{eq:paths}). Then
\begin{equation}\label{eq:frames}
f_\alpha^{(2)}(t,s)-f_\alpha^{(1)}(t,s) = st\,\big( P_{\dot\gamma}^\perp (R(\dot\gamma,\gamma')e_\alpha)+\Delta\Lambda^\beta{}_\alpha e_\beta \big)_{|(0,0)}+\mathcal{O}(t^2,(s,t)^3)\,,
\end{equation}
where $P_{\dot\gamma}^\perp = \delta + n_{\dot \gamma}\otimes g(n_{\dot\gamma},\cdot)$ is the projection orthogonal to $\dot\gamma$. The term $\Delta\Lambda$ is sourced by the relative velocity $v=\nabla_{\dot\gamma}\gamma'$ and acceleration $a=\nabla_{\dot\gamma}\nabla_{\dot\gamma}\gamma'$; with $N=1-g(\dot\gamma,n_{\gamma'})$,
\begin{equation}\label{eq:dL}
\Delta\Lambda^\beta{}_\alpha{}_{|(0,0)}=\frac{1}{N}\Big[g(n_{\gamma'},e_\alpha)a^\beta-g(a,e_\alpha)n_{\gamma'}^\beta+\frac{1}{N}\frac{g(v,n_{\gamma'}+\dot\gamma)}{\sqrt{|g(\gamma',\gamma')}|}\big(g(n_{\gamma'},e_\alpha)v^\beta-g(v,e_\alpha)n_{\gamma'}^\beta\big)\Big]_{|(0,0)}\,.
\end{equation}}

We present the details of the rather lengthy proof of this theorem in appendix \ref{sec:proof}. As mentioned before, this theorem generalizes the well-known result on the path-dependence of parallel transport, see section \ref{sec:tp}. Here the deviation of the frames $f_\alpha^{(1)}$ and $f_\alpha^{(2)}$ due to the path dependence of Fermi--Walker transport is interpreted as the desynchronization of two gyroscope systems. This is caused on the one hand by the Riemann curvature tensor of spacetime, and on the other hand by contributions $\Delta\Lambda$ from Lorentz boosts that still appear on flat spacetime. The contributions from the Rimeann tensor are related to the so called Lense-Thirring and geodetic effect, while those from the Lorentz boosts are related to the Thomas precession and capture the simple fact that the second observer in general is boosted and rotates relatively to the first. There is no Thomas precession in case that the spatial part of the relative velocity and acceleration stay parallel to the spatial initial direction~$\gamma'$ of separation between the two observers. 

\subsection{Local rotation theorem}\label{sec:localrot}
We will now reformulate the result of Theorem~1 on the desynchronization of gyroscope systems into a statement on their relative rotation induced by the gravitational field. This can also be interpreted as the relative rotation of two nearby observers whose spatial frames coincide with the respective gyroscope systems.

In order to do so, a first observer prepares two aligned sets of three gyroscopes with orthonormal spin axes, representing her spatial frame. One of the gyroscope systems is passed to a second nearby observer via free-fall. When received by the second observer it is still orthonormal, and so can be used to define the second observer's spatial frame. As a consequence of this procedure, both observers consider their spatial frames to be aligned. After some time the first observer passes the remaining set of gyroscopes to the second who compares the alignment of both gyroscope systems and deduces a relative rotation of the spatial frames from their desynchronization. The result is summarized in the following theorem:

\vspace{6pt}\noindent\textbf{Theorem 2.} \textit{Consider a freely falling observer ${}^{(1)}$ with orthonormal frame $\{e_\mu\}$ at some point $p$ of a metric manifold $(M,g)$, and a general second observer ${}^{(2)}$ separated from the first by parameter distance $s$ along a timelike geodesic through $p$ with tangent $X$. Then the gravitational field and the observers' relative motion induce a relative spatial frame rotation $\Delta\Omega=\Omega^{(2)}-\Omega^{(1)}$ given by
\begin{equation}\label{eq:t2}
\Delta\Omega^{\beta}{}_\alpha (s)e_\beta  = s\Big( P_{e_0}^\perp (R(e_0,X)e_\alpha)+\Xi^\beta{}_\alpha e_\beta \Big)+\mathcal{O}(s^2)\,,
\end{equation}
where, writing $M=1-g(e_0,n_X)$, and $v$ and $a$ for the observers' relative velocity and acceleration,
\begin{equation}\label{eqn:Xi}
\Xi^\beta{}_\alpha = \frac{1}{M}\Big[g(n_X,e_\alpha)a^\beta-g(a,e_\alpha)n_X^\beta+\frac{1}{M}\frac{g(v,n_X+e_0)}{\sqrt{|g(X,X)}|}\big(g(n_X,e_\alpha)v^\beta-g(v,e_\alpha)n_X^\beta\big)\Big].
\end{equation}}

Before we prove this theorem, we recall that a general observer on a worldline with unit normalized four-velocity $e_0$ and orthonormal spatial axes $e_\alpha$ (defined as vector fields along the worldline) is described by the equations
\begin{equation}\label{eq:rot}
\nabla_{e_0}e_0 = A^\alpha e_\alpha \,,\quad \nabla_{e_0} e_\alpha =  A_\alpha e_0+\Omega^\beta{}_\alpha e_\beta\,. 
\end{equation}
The first equation describes the covariant acceleration and turns into geodesic motion for $A^\alpha=0$. The second equation, for $\Omega\neq 0$, describes the rotation of the spatial frame in time.

\vspace{6pt}\noindent \textit{Proof of Theorem 2.} The measurement of the relative rotation of two observers is performed by comparing spatial frames of reference that are modelled by gyroscope systems as described in Theorem~1. Hence we identify the frame vectors $e_\mu\equiv e_\mu(0,0)$, where $e_0(0,0) = \dot\gamma(0,0)$, and $X\equiv \gamma'(0,0)$. We begin by applying the covariant time-derivative $\nabla_{n_{\dot\gamma}}$ at $t=0$ to equation~(\ref{eq:frames}), which yields
\begin{equation}\label{eq:lefthand}
\nabla_{n_{\dot\gamma}}f_\alpha^{(2)}(0,s)-\nabla_{n_{\dot\gamma}}f_\alpha^{(1)}(0,s) = s\,\big( P_{\dot\gamma}^\perp (R(\dot\gamma,\gamma')e_\alpha)+\Delta\Lambda^\beta{}_\alpha e_\beta \big)_{|(0,0)}+\mathcal{O}(s^2)\,.
\end{equation}
Since $n_{\dot\gamma}\equiv f_0^{(1,2)}$ completes both spatial frames $f_\alpha^{(1,2)}$ into complete orthonormal frames, the left hand side can be written as in equation~(\ref{eq:rot}),
\begin{equation}\label{eq:1111}
\Big(A^{(2)}_\alpha(0,s)- A^{(1)}_\alpha(0,s)\Big) n_{\dot\gamma}(0,s)+\Omega^{(2)\,\beta}{}_\alpha(0,s) f_\beta^{(2)}(0,s)-\Omega^{(1)\,\beta}{}_\alpha(0,s) f_\beta^{(1)}(0,s)\,,
\end{equation}
where $\Omega^{(2)}$ describes the frame rotation of the second observer, while $\Omega^{(1)}$ describes the frame rotation of the first observer as seen from the perspective of the second. Both gyroscope systems $f^{(1)}$ and $f^{(2)}$ move along the same worldline, so that $\nabla_{n_{\dot \gamma}}n_{\dot\gamma}=A^{(1)\alpha}f^{(1)}_\alpha=A^{(2)\alpha}f^{(2)}_\alpha$; then, since $f^{(1)}_\alpha(0,s)=f^{(2)}_\alpha(0,s)$, the covariant acceleration terms above vanish. In consequence, expression~(\ref{eq:1111}) simplifies to
\begin{equation}\label{eq:2222}
\big(\Omega^{(2)\,\beta}{}_\alpha(0,s)-\Omega^{(1)\,\beta}{}_\alpha(0,s)\big) f_\beta^{(2)}(0,s)\,.
\end{equation}
Now observe that $\Omega^{(1,2)}(0,s)\sim\mathcal{O}(s)$. To see this, we can rewrite 
\begin{eqnarray}
A^{(1,2)}_\alpha(0,0)e_0(0,0) + \Omega^{(1,2)}{}^\beta{}_\alpha(0,0)e_\beta(0,0) & = & \nabla_{n_{\dot\gamma}}f_\alpha^{(1,2)}(0,0) \\
&= & \big(\nabla_{n_{\dot\gamma}}f_\alpha^{(1,2)}(t,0)\big)_{|t=0}=\big(\nabla_{n_{\dot\gamma}}e_\alpha(t,0)\big)_{|t=0}=0 \nonumber
\end{eqnarray}
since the derivative does not act on the $s$-dependence, since by construction $f_\alpha^{(1,2)}(t,0)=e_\alpha(t,0)$ because the combined finite Lorentz boosts become trivial, and since the $e_\alpha(t,0)$ are parallelly transported. This indeed implies the condition of a vanishing zeroth order $\Omega^{(1,2) \beta}{}_\alpha(0,0)=0$.
Hence we can replace $f_\alpha^{(2)}(0,s)$ in (\ref{eq:2222}) by its zeroth order expressions $f_\alpha^{(2)}(0,0)=e_\alpha(0,0)$. This finally yields
\begin{eqnarray}
\big(\Omega^{(2)\,\beta}{}_\alpha(0,s) -\Omega^{(1)\,\beta}{}_\alpha(0,s)\big) e_\beta(0,0) = \Delta\Omega^\beta{}_\alpha(s)e_\beta(0,0)
\end{eqnarray}
for the left hand side of (\ref{eq:lefthand}). The components $\Xi^\beta{}_\alpha$ arise from the $\Delta\Lambda^\beta{}_\alpha$ by the frame vector identifications made above. This completes the proof. $\square$

\vspace{6pt}In order to give a precise description of a spatial frame of reference, observers need to have a stable notion of spatial axes. Experimentally, this can be realized by means of gyroscope systems. Theorem~2 tells us the interesting fact, that observers who align their spatial frame at some time will lose this synchronization due to gravitational effects and relative motion effects. In order to uphold aligned spatial frames in spacetime, observers have to counterbalance these effects continually. 

One also observes that the gravitational effects on the rotation of spatial frames of reference are sourced by the magnetic $R^\beta{}_{\alpha 0\delta}$ components  of the Riemann tensor with respect to the first observer's frame, since $P_{\dot\gamma}^\perp (R(\dot\gamma,\gamma')e_\alpha)=R^\beta{}_{\alpha 0\delta}\gamma'{}^\delta \,e_\beta$. This provides a nice picture of the Riemann tensor when combined with the Jacobi-equation which tells us that the $R^\alpha{}_{0 \beta 0}$ components of the Riemann tensor are responsible for the relative acceleration between nearby observers.

\section{Illustration: Schwarzschild or Kerr spacetime?}\label{sec:kerr}
In this section we will illustrate our results for Kerr and Schwarzschild spacetime. We propose a very simple local experiment which enables observers to determine which of these backgrounds describes their central mass, or, intuitively, whether their spacetime has angular momentum or not. More precisely, we will demonstrate that there exists a class of observers on Kerr spacetime which see an additional relative rotation of nearby gyroscope systems caused by the angular momentum parameter in the metric, while  the analogue class of observers on Schwarzschild spacetime does not observe this effect. 

We follow the notation from \cite{Muller:2009bw} in this section.
Consider the metric $g$ of Kerr spacetime in Boyer-Lindquist coordinates $(t,r,\theta,\phi)$,
\begin{equation}
g=-\big(1-\frac{r_s r}{\Sigma}\big)dt^2-\frac{2 r_s a r \sin^2\theta}{\Sigma}dtd\phi+\frac{\Sigma}{\Delta}dr^2+\Sigma d\theta^2+\big(r^2+a^2+\frac{r_s a^2 r \sin^2\theta}{\Sigma}\big)\sin^2\theta d\phi^2
\end{equation}
with $\Sigma=r^2+a^2 \cos^2\theta$ and $\Delta=r^2-r_s r+a^2$, where $r_s$ is the Schwarzschild radius and $a$ the angular momentum parameter. Moreover, consider a general stationary observer on Kerr spacetime whose frame $\{e_\mu\}$ is given by 
\begin{equation}
e_0=\Gamma(\partial_t+\Omega\partial_\phi),\quad e_1=\sqrt{\frac{\Delta}{\Sigma}}\partial_r,\quad e_2=\frac{1}{\sqrt{\Sigma}}\partial_\theta,\quad e_3=\frac{\Gamma}{\sqrt{\Delta}\sin\theta}\big((g_{t\phi}+\Omega g_{\phi\phi})\partial_t-(g_{tt}+\Omega g_{t\phi})\partial_\phi\big)\,,
\end{equation}
where $\Gamma=\sqrt{-g_{tt}-2\Omega g_{t\phi}-\Omega^2 g_{\phi\phi}}$ and the parameter $\Omega$ is the observer's angular velocity.

According to Theorem 2, the spatial frames of nearby observers, or nearby spatial gyroscope systems, rotate relatively to each other due to the non-vanishing $R^\beta{}_{\alpha\gamma 0}$ components of the Riemann tensor. We now calculate this tensor in the chosen stationary frame; it can best be expressed in Petrov index notation,  where the antisymmetry in the index pairs $[\mu\nu],\,[\rho\sigma]$ and the exchange symmetry of these pairs in $R_{\mu\nu\rho\sigma}$ are used to identify all Riemann tensor components with a symmetric $6\ \times\ 6$ matrix. The six-indices are ordered as $[01],[02],[03],[12],[13],[23]$. In our case this matrix takes the form
\begin{equation}\label{eqn:inv2}
R_{[\mu\nu][\rho\sigma]} \sim \left(\begin{matrix}
A& B & 0 & 0 & C &D\\
 B & F & 0 & 0 & E & -C\\
 0 & 0 & G & H & 0  & 0\\
 0 & 0 & H & -G & 0 & 0\\
 C & E & 0 & 0 & -F &B\\
 D & -C & 0 & 0 & B & -A
\end{matrix}\right)\!,
\end{equation}
in terms of rather involved functions $A,B,C,D,E,F,G$ and $H$ that depend on the coordinates $r,\theta$ and the parameters $r_s, a$ and $\Omega$. Below the only relevant functions will be $C$ and $D$. Using some short-hand notation displayed in appendix \ref{app:kerr}, their structural form is
\begin{eqnarray}
C&=&\frac{3r_s r \sin\theta \sqrt{a^2 + r (-r_s + r)}C_{0}(1+C_{1}\Omega+C_2\Omega^2)}{32 \Sigma^5(g_{tt}+2 g_{t\phi}\Omega+g_{\phi\phi}\Omega^2)},\nonumber\\
D&=&\frac{a r_s \cos\theta D_0(D_1+D_2\Omega+D_3\Omega^2)}{32\sqrt{2}\Sigma^5(g_{tt}+2g_{t\phi}\Omega+g_{\phi\phi}\Omega^2)}\,.
\end{eqnarray}

We may now imagine the following simple experiment: Consider two stationary observers, the first moving along a geodesic and the second moving along some wordline at fixed spatial distance in the radial $e_1$-direction. These observers could be realized either via two satellites, or by one geodetically moving satellite and a second observer on the surface of the central mass. Now the two observers perform the gyroscope transport experiment. The geodesic observer prepares two systems of gyroscopes which are aligned at her position. One of these systems is sent to the second observer immediately, the other system after some time has passed. Then the second observer compares the two different gyroscope systems. Their relative rotation is given by formula~(\ref{eq:t2}) in Theorem~2. This formula simplifies considerably because the two different observers are at fixed distance; hence their relative velocity and relative acceleration vanishes. Moreover, the timelike direction $X$ along which the gyroscopes were transported between the observers has only radial spatial components. Hence, 
\begin{equation}
\Delta\Omega^{\beta}{}_\alpha (s)e_\beta  = s\, P_{e_0}^\perp (R(e_0,X)e_\alpha) + \mathcal{O}(s^2) = s\, R^\beta{}_{\alpha 0 1}X^1 \,e_\beta + \mathcal{O}(s^2)\,,
\end{equation}  
where the relevant curvature components can be read off from equation (\ref{eqn:inv2}). This yields
\begin{equation}
\Delta\Omega^{\beta}{}_\alpha (s) = s \left(\begin{array}{ccc} 0 & 0 & C \\ 0 & 0 & D\\ -C & -D & 0 \end{array} \right)X^1 +\mathcal{O}(s^2)\,.
\end{equation}
This result has the following effects. Gyroscopes initially synchronized along the $e_2$-direction experience a desynchronization by
\begin{equation}\label{eq:LTeffect}
\Delta\Omega^{\beta}{}_2 (s)e_\beta= s D X^1 e_3\,,
\end{equation}
while gyroscopes initially synchronized along $e_1$ are subject to a relative change due to
\begin{equation}\label{eq:geod}
\Delta\Omega^{\beta}{}_1 (s)e_\beta= s C X^1 e_3\,.
\end{equation}

Now observe that in the zero angular momentum limit $a\rightarrow 0$ the function $D$ vanishes but $C$ does not. This means there occurs no desynchronization for gyroscopes initially pointing into the $e_2$-direction in Schwarzschild spacetime. This identifies equations (\ref{eq:LTeffect}) and (\ref{eq:geod}) respectively as non-perturbative covariant versions of the Lense-Thirring effect and the geodetic effect~\cite{Schiff}. Thus, using gyroscopes initially pointing in the azimuthal direction, stationary observers can answer the question, whether their spacetime is rotating or not. Crucially, this can be achieved by a local experiment without the need of global information at infinity, or spacetime perturbation theory.

\section{Discussion}\label{sec:disc}
In this article we have carefully analyzed the effects of curved spacetime on gyroscope motion. We investigated how initially aligned gyroscope systems moving relatively to each other  through spacetime lose their synchronization. As our central result, we have proven Theorem~1 which makes it clear that this desynchronization is caused by the magnetic components of the Riemann curvature tensor and by the relative motion of the gyroscope systems. In order to derive this result we studied the path dependence of Fermi--Walker transport: spin axes were transported along different paths between the same initial and final points in spacetime, taking care of subtle non-differentiability issues. Our results  extend the well-known theorem about the path-dependence of parallel transport. A direct physical consequence of Theorem~1 is the relative rotation of spatial observer frames as we have derived in Theorem~2. This result shows that even if two observers at a fixed distance in a laboratory initially align their spatial frames, they will in general lose this synchronization. In order to keep their spatial frames synchronized, they must counterbalance a rotation induced by the curvature of spacetime.

In general, equation (\ref{eq:t2}) of Theorem~2 catches in a fully covariant way and without involving spacetime perturbations the geodetic and frame dragging effects as well as the Thomas precession. The first two effects are encoded in the curvature contributions, while the Thomas precession, as a special relativistic effect, appears in the velocity and acceleration dependent second term, see~(\ref{eqn:Xi}). A comparison to standard formulae for these effects is desirable; it would require a detailed analysis of the different experimental setups used for the derivations and, moreover, a rewriting of our covariant expressions in terms of a suitable time-space split and a post-Newtonian approximation. 

As an application of our new results we proposed a local gyroscope experiment which enables stationary observers to decide whether their spacetime possesses angular momentum in the sense of the Kerr metric or not.  If this experiment was realized, it could support the data about frame dragging and the geodetic effect collected by the Gravity Probe B~\cite{Everitt:2011hp}.

\acknowledgments CP and MNRW thank Martin Schasny, Manuel Hohmann, Matthias Lange, Falk Lindner, Gunnar Preiss and Felix Tennie for inspiring discussions. They acknowledge partial financial support from the German Research Foundation DFG under grant WO 1447/1-1.

\appendix
\section{Proof of Theorem 1}\label{sec:proof}
The proof of Theorem~1 in section~\ref{sec:ex} involves the comparison of spatial orthonormal gyroscope systems transported along the two paths~${}^{(1)}$ and~${}^{(2)}$ of a closed infinitesimal rectangle. The sequence~(\ref{eq:paths}) and figure~\ref{fig:ex} describe in detail how this transport is accomplished. We will now work through the steps of the two different paths. First, in section~\ref{sec:proofstruc}, we express the difference of the final gyroscope axes by the Riemann tensor and a second term that acts as a further transformation on the initial gyroscope axes. Second, in~\ref{sec:proofdet}, we will calculate this transformation, which depends on the relative velocity and acceleration of the gyroscope systems, to first nontrivial order. 

The notational conventions here are those introduced in section \ref{sec:basicgm}. We use the abbreviation $n_X=\frac{X}{\sqrt{|g(X,X)|}}$ for the normalized vector in direction $X$; indices $\alpha, \beta, \gamma, \dots $ run from $1, \dots, 3$, while indices $\mu,\nu,\rho, \dots $ run from $0, \dots, 3$.

\subsection{Structure of the result}\label{sec:proofstruc}

We begin by calculating the transport of the initial gyroscope system $e_\alpha(0,0)$ along the path
\begin{equation}
{}^{(1)}: e_\alpha(0,0) \overset{\dot\gamma(t,0)}{\longrightarrow} e_\alpha(t,0) \overset{\Lambda_{11}(t)}{\curvearrowright} e_\alpha^{(1)}(t,0) \overset{\gamma'(t,s)}{\longrightarrow} e_\alpha^{(1)}(t,s)\overset{\Lambda_{12}(t,s)}{\curvearrowright} f_\alpha^{(1)}(t,s)\,.
\end{equation}
At every stage of the transport we complete the spatial gyroscope system by the normalized timelike tangent along which it is transported into a full orthonormal frame. In this way the $\{e^{(1)}_\alpha(t,s)\}$ form a complete frame $\{e^{(1)}_\mu(t,s)\}$ by setting $e^{(1)}_0(t,s)=n_{\gamma'}(t,s)$; similarly the $\{e_\alpha(t,0)\}$ form a complete frame $\{e_\mu(t,0)\}$ by setting $e_0(t,0)=n_{\dot\gamma}(t,0)=\dot\gamma(t,0)$. The final axes $f^{(1)}_{\alpha}(t,s)$ of the gyroscope system are constructed with the help of the Lorentz transformation $\Lambda_{12}(t,s)$ from $e^{(1)}_{\mu}(t,s)$. These in turn are constructed from the $e^{(1)}_{\mu}(t,0)$ by gyroscope transport along $\gamma'(t,s)$, which at this point simply is parallel transport since we assumed free fall for the transport of the gyroscopes between the observers.  Hence
\begin{equation}
f^{(1)}_{\alpha}(t,s)=\Lambda_{12}{}^\mu{}_\alpha(t,s) e^{(1)}_\mu{}^\rho(t,0)p^{(1)}_\rho(t,s)
\end{equation}
expressed with respect to the basis $p^{(1)}_\mu(t,s)$ constructed by parallel transported of the frame $e_\mu(0,0)$ along path~${}^{(1)}$. Here we used the fact that the components of a parallelly transported vector expressed with respect to a parallelly transported basis do not change along the path. Now observe that the $e^{(1)}_\mu(t,0)$ are constructed from the $e_\mu(t,0)$ by the Lorentz transformation $\Lambda_{11}(t)$, and the $e_\mu(t,0)$ are given by parallel transport of the initial frame $e_{\mu}(0,0)$. Using the parallelly transported basis $p^{(1)}_\mu$ and the fact that with respect to this basis $e_\mu{}^\nu(0,0)=\delta^\nu_\mu$ we obtain the following result for the gyroscope system transport along path ${}^{(1)}$:
\begin{equation}
f^{(1)}_{\alpha}(t,s)=\Lambda_{12}{}^\mu{}_\alpha(t,s)\Lambda_{11}{}^\rho{}_\mu(t) e_\rho{}^\sigma(0,0)p^{(1)}_\sigma(t,s)
=\Lambda_{12}{}^\mu{}_\alpha(t,s)\Lambda_{11}{}^\rho{}_\mu(t) p^{(1)}_\rho(t,s)\,.
\end{equation}

We now analyze the gyroscope system transport along the path
\begin{equation}
{}^{(2)}: e_\alpha(0,0) \overset{\Lambda{21}}{\curvearrowright} e_\alpha^{(2)}(0,0) \overset{\gamma'(0,s)}{\longrightarrow} e_\alpha^{(2)}(0,s) \overset{\Lambda_{22}(0,s)}{\curvearrowright} f_\alpha^{(2)}(0,s) \overset{\dot\gamma(t,s)}{\longrightarrow} f_\alpha^{(2)}(t,s)\,.
\end{equation}
As above the initial spatial gyroscope axes $e_\alpha(0,0)$ can be completed into a full frame by means of the normalized tangent of the path they are transported along. The final gyroscope axes $f_\alpha^{(2)}(t,s)$ after transport along path~${}^{(2)}$ are constructed from the $f_\alpha^{(2)}(0,s)$ by gyroscope transport. In general this is not a parallel transport but Fermi--Walker transport since we did not make any assumptions about the path $\gamma(t,s)$ at fixed $s\neq 0$. Introducing the basis $p^{(2)}_\mu(s,t)$ constructed by parallel transport of the initial frame $e_\mu(0,0)$ along path ${}^{(2)}$ we may write
\begin{eqnarray}
f_\alpha^{(2)}(t,s) &=& f^{(2)}_\alpha{}^\rho(0,s)p^{(2)}_\rho(t,s)+t\,\nabla_{\dot\gamma} f^{(2)}_\alpha{}^\rho(0,s)\, p^{(2)}_\rho(t,s)+\mathcal{O}(t^2) \nonumber\\
&=& f^{(2)}_\alpha{}^\rho(0,s)p^{(2)}_\rho(t,s)+t\big[g(\nabla_{\dot\gamma}n_{\dot\gamma}, f^{(2)}_\alpha)n_{\dot\gamma}{}^\rho\big]_{|(0,s)}p^{(2)}_\rho(t,s)+\mathcal{O}(t^2)
\end{eqnarray} 
by using the covariant Taylor series as in~(\ref{eq:covt}) and the gyroscope transport equation~(\ref{eq:gyro}). Tracing back the $f^{(2)}_\alpha{}^\rho(0,s)$ to the frame $e_\mu(0,0)$ works similarly as for path $(1)$, since they are built from Lorentz transformations and parallel transport. So we obtain: 
\begin{equation}
f_\alpha^{(2)}(t,s)= \Lambda_{22}{}^\sigma{}_\alpha(s)\Lambda_{21}{}^\rho{}_\sigma p^{(2)}_\rho(t,s)+t\big[g(\nabla_{\dot\gamma}n_{\dot\gamma}, f^{(2)}_\alpha)n_{\dot\gamma}{}^\rho\big]_{|(0,s)}p^{(2)}_\rho(t,s)+\mathcal{O}(t^2)\,.
\end{equation} 
To simplify further we expand the components of the second term to first order in~$s$ by using the parallelly transported orthonormal basis $p^{(2)}_\mu$, the covariant Taylor series and $\nabla_{\dot\gamma}\dot\gamma_{|(0,0)}=0$, which leads to
\begin{eqnarray} 
g(\nabla_{\dot\gamma}n_{\dot\gamma}, f^{(2)}_\alpha)_{|(0,s)}&=&\eta_{\mu\nu}\big(\nabla_{\dot\gamma}n_{\dot\gamma}\big)^\mu{}_{|(0,s)} f^{(2)}_\alpha{}^\nu(0,s)\nonumber\\
&=& \eta_{\mu\nu}s\big(\nabla_{\gamma'}\nabla_{\dot\gamma}n_{\dot\gamma}\big)^\mu{}_{|(0,0)}\Lambda_{22}{}^\sigma{}_\alpha(0)\Lambda_{21}{}^\nu{}_\sigma+\mathcal{O}(s^2)\nonumber\\
&=&s \, g(\nabla_{\gamma'}\nabla_{\dot\gamma}n_{\dot\gamma},e_\alpha)_{|(0,0)}+\mathcal{O}(s^2)\,.
\end{eqnarray}
The last equality is due to order counting and the fact that $\Lambda_{22}{}^\sigma{}_\alpha(0)\Lambda_{21}{}^\nu{}_\sigma=\delta^\nu_\alpha$ since both Lorentz boosts cancel at the origin. Using $n_{\dot\gamma}{}^\rho{}_{|(0,s)}=\dot\gamma^\rho{}_{|(0,0)}+\mathcal{O}(s)$, we collect the following result for the gyroscope system transport along path ${}^{(2)}$:
\begin{equation}
f_\alpha^{(2)}(t,s)=\Lambda_{22}{}^\sigma{}_\alpha(s)\Lambda_{21}{}^\rho{}_\sigma p^{(2)}_\rho(t,s)+ st \big[g(\nabla_{\gamma'}\nabla_{\dot\gamma}n_{\dot\gamma},e_\alpha)\dot\gamma \big]_{|(0,0)}+\mathcal{O}(t^2)+\mathcal{O}((t,s)^3)\,.
\end{equation}

We now calculate the difference between the axes of the gyroscope systems transported along the two different paths. We use the standard theorem for the path-dependence of parallel transport~(\ref{eq:pd}) to write $p^{(1)}_\rho(t,s)=p^{(2)}_\rho(t,s)-st \,R(\dot\gamma, \gamma')e_\rho{}_{|(0,0)}+\mathcal{O}((t,s)^3)$, and that $\Lambda_{12}{}^\mu{}_\alpha(0,0)\Lambda_{11}{}^\rho{}_\mu(0)=\delta^\rho_\alpha$. Hence
\begin{eqnarray}\label{eq:gyrodiff}
f_\alpha^{(2)}(t,s)-f_\alpha^{(1)}(t,s)&=&\big(\Lambda_{22}{}^\sigma{}_\alpha(s)\Lambda_{21}{}^\rho{}_\sigma-\Lambda_{12}{}^\mu{}_\alpha(t,s)\Lambda_{11}{}^\rho{}_\mu(t)\big) p^{(2)}_\rho(t,s)\nonumber \\
&&+st \big[R(\dot\gamma,\gamma')e_\alpha + g(\nabla_{\gamma'}\nabla_{\dot\gamma}n_{\dot\gamma},e_\alpha)\dot\gamma \big]_{|(0,0)}+\mathcal{O}(t^2,(t,s)^3)\,.
\end{eqnarray}
Note that also the first line of this result is of $\mathcal{O}(st)$. Indeed, we already know that $\Lambda_{22}{}^\sigma{}_{\alpha}(0)\Lambda_{21}{}^\nu{}_{\sigma}=\delta^\nu_\alpha$ and $\Lambda_{12}{}^\sigma{}_{\alpha}(0,0)\Lambda_{11}{}^\nu{}_{\sigma}(0)=\delta^\nu_\alpha$; furthermore $\Lambda_{12}{}^\sigma{}_{\alpha}(t,0)\Lambda_{11}{}^\nu{}_{\sigma}(t)=\delta^\nu_\alpha$ since these transformations act at the same point of the transport, and $\Lambda_{12}{}^\sigma{}_{\alpha}(0,s)\Lambda_{11}{}^\nu{}_{\sigma}(0)=\Lambda_{22}{}^\sigma{}_{\alpha}(s)\Lambda_{21}{}^\nu{}_{\sigma}$ since both paths are identical for $t=0$; hence we may write
\begin{equation}\label{eq:DL}
\Lambda_{22}{}^\sigma{}_{\alpha}(s)\Lambda_{21}{}^\rho{}_{\sigma}-\Lambda_{12}{}^\mu{}_{\alpha}(t,s)\Lambda_{11}{}^\rho{}_{\mu}(t)=st\, \Delta \Lambda^\rho{}_\alpha{}_{|(0,0)}+\mathcal{O}((s,t)^3)\,. 
\end{equation}
Inserting this into equation (\ref{eq:gyrodiff}) yields
\begin{eqnarray}\label{eq:diffF}
f_\alpha^{(2)}(t,s)-f_\alpha^{(1)}(t,s) &=& st \big[R(\dot\gamma,\gamma')e_\alpha +\Delta\Lambda^\rho{}_\alpha e_\rho+ g(\nabla_{\gamma'}\nabla_{\dot\gamma}n_{\dot\gamma},e_\alpha)\dot\gamma \big]_{|(0,0)}+\mathcal{O}(t^2,(t,s)^3)\\
&=&  st \big[P_{\dot\gamma}^\perp(R(\dot\gamma,\gamma')e_\alpha) + \Delta\Lambda^\rho{}_\alpha e_\rho+  g(\nabla_{\dot\gamma}\nabla_{\gamma'}n_{\dot\gamma},e_\alpha)\dot\gamma \big]_{|(0,0)}+\mathcal{O}(t^2,(t,s)^3)
\,,\nonumber
\end{eqnarray}
where $P_{\dot\gamma}^\perp = \delta+ n_{\dot\gamma}\otimes g(n_{\dot\gamma},\cdot)$ denotes the projection orthogonal to $\dot \gamma$. 

Observe that the expression derived above already has the structure found in Theorem~1.

\subsection{Detailed calculation}\label{sec:proofdet}
Two points remain to be proven: we need to show first that $\Delta\Lambda^0{}_\alpha{}_{|(0,0)} = -g(\nabla_{\dot\gamma}\nabla_{\gamma'}n_{\dot\gamma},e_\alpha)_{|(0,0)}$ so that the third term in~(\ref{eq:diffF}) is cancelled, and second that the spatial components $\Delta\Lambda^\beta{}_\alpha{}_{|(0,0)}$ have the form claimed in~(\ref{eq:dL}).
 
To extract $\Delta\Lambda^\rho{}_\alpha{}_{|(0,0)}$ from~(\ref{eq:DL}) we calculate the $\partial_s\partial_t$ derivative and evaluate it at $(t,s)=(0,0)$:
\begin{equation}
\Delta\Lambda^\rho{}_\alpha{}_{|(0,0)}=-\partial_s\partial_t(\Lambda_{12}{}^\mu{}_\alpha(t,s))_{|(0,0)}\Lambda_{11}{}^\rho{}_{\mu}(0)-\partial_s(\Lambda_{12}{}^\mu{}_{\alpha}(t,s))_{|(0,0)}\partial_t(\Lambda_{11}{}^\rho{}_{\mu}(t))_{|0}\,.
\end{equation}
The Lorentz transformations of interest are fixed by the relations
\begin{equation}
f^{(1)}_{\mu}(t,s)=\Lambda_{12}{}^\rho{}_{\mu}(t,s)e^{(1)}_{\rho}(t,s),\quad e^{(1)}_{\mu}(t,0)=\Lambda_{11}{}^\rho{}_{\mu}(t)e_{\rho}(t,0)\,,
\end{equation}
and by their action on the spatial frame vectors as in equation~(\ref{eq:boost}). With the abbreviation $N=1-g(n_{\dot\gamma},n_{\gamma'})$ we find
\begin{equation}
\Lambda_{12}{}^0{}_{\alpha}(t,s) = g(n_{\dot\gamma}, e^{(1)}_\alpha)_{|(t,s)},\quad 
\Lambda_{12}{}^\beta{}_{\alpha}(t,s)=\delta^\beta_\alpha+\Big(\frac{\delta^{\beta\sigma}g(n_{\dot\gamma}, e^{(1)}_\sigma)g(n_{\dot\gamma}, e^{(1)}_\alpha)}{N}\Big)_{|(t,s)}
\end{equation}
and
\begin{equation}
\Lambda_{11}{}^\rho{}_{0}(t)  = \eta^{\rho\mu} g(n_{\gamma'}, e_\mu)_{|(t,0)}\,,\quad 
\Lambda_{11}{}^0{}_{\alpha}(t)= g(n_{\gamma'},e_\alpha)_{|(t,0)}\,,
\end{equation}
\begin{equation}
\Lambda_{11}{}^\beta{}_{\alpha}(t)=\delta^\beta_\alpha+\Big(\frac{\delta^{\beta\sigma}g(n_{\gamma'}, e_\sigma)g(n_{\gamma'}, e_\alpha)}{N}\Big)_{|(t,0)}\,.
\end{equation}

It is now convenient to employ the following relations for the covariant derivatives of the normalized tangent vectors 
\begin{eqnarray}
 (\nabla_{\gamma'}n_{\dot\gamma})_{|(0,0)}&=&(\nabla_{\gamma'}\dot\gamma)_{|(0,0)}+(g(\nabla_{\gamma'}\dot\gamma,\dot\gamma)\dot\gamma)_{|(0,0)}\\
(\nabla_{\dot\gamma}n_{\gamma'})_{|(0,0)}&=&\Big(\frac{\nabla_{\dot\gamma}\gamma'}{\sqrt{|g(\gamma',\gamma')|}}\Big)_{|(0,0)}+\Big(\frac{g(\nabla_{\dot\gamma}\gamma',n_{\gamma'})n_{\gamma'}}{\sqrt{|g(\gamma',\gamma')|}}\Big)_{|(0,0)}
\end{eqnarray} 
in order to equate the required components of the Lorentz transformations and their derivatives:
\begin{eqnarray}
&&\Lambda_{11}{}^0{}_{0}(0) = \eta^{0\mu}g(n_{\gamma'}, e_\mu)_{|(0,0)}\,,\qquad
\Lambda_{11}{}^0{}_{\beta}(0) = g(n_{\gamma'},e_\beta)_{|(0,0)},\nonumber\\
&&\Lambda_{11}{}^\delta{}_{0}(0) = \eta^{\delta\mu}g(n_{\gamma'}, e_\mu)_{|(0,0)}\,,\qquad
\Lambda_{11}^\delta{}_{\beta}(0) = \delta^\delta_\beta+\Big(\frac{\delta^{\delta\epsilon}g(n_{\gamma'}, e_\epsilon)g(n_{\gamma'}, e_\beta)}{N}\Big)_{|(0,0)}\,,
\end{eqnarray}
\begin{eqnarray}
&& \partial_t\Lambda_{11}{}^0{}_{0}{}_{|(0)} = \eta^{0\mu}g(\nabla_{\dot\gamma} n_{\gamma'}, e_\mu)_{|(0,0)}\,,\qquad
\partial_t\Lambda_{11}{}^0{}_{\beta}{}_{|(0)} = g(\nabla_{\dot\gamma}n_{\gamma'},e_\alpha)_{|(0,0)}\,,\nonumber\\
&&\partial_t\Lambda_{11}{}^\delta{}_{0}{}_{|(0)} = \eta^{\delta\mu}g(\nabla_{\dot\gamma}n_{\gamma'}, e_\mu)_{|(0,0)}\,,\\
&&\partial_t\Lambda_{11}{}^\delta{}_{\beta}{}_{|(0)} = \Big[\frac{1}{2 N}\delta^{\delta\epsilon}g(\nabla_{\dot\gamma}n_{\gamma'},e_{(\epsilon})g(e_{\beta)},n_{\gamma'})+\frac{g(\nabla_{\dot\gamma}n_{\gamma'},n_{\dot\gamma})}{N^2}\delta^{\delta\epsilon}g(n_{\gamma'},e_\beta)g(n_{\gamma'},e_\epsilon)\Big]_{|(0,0)}\,,\nonumber
\end{eqnarray}
\begin{eqnarray}
&& \partial_s\Lambda_{12}{}^0{}_{\alpha}{}_{|(0,0)}=\Big[g(\nabla_{\gamma'}n_{\dot\gamma},e_\alpha)+\frac{1}{N}g(n_{\gamma'},e_\alpha)g(\nabla_{\gamma'}n_{\dot\gamma},n_{\gamma'})\Big]_{|(0,0)}\,,\\
&&\partial_s\Lambda_{12}{}^\beta{}_{\alpha}{}_{|(0,0)}=-\delta^{\epsilon\beta}\Big[\frac{2}{N}g(\nabla_{\gamma'}n_{\dot\gamma},e_{(\epsilon})g(e_{\alpha)},n_{\gamma'})+\frac{1}{N^2}g(\nabla_{\gamma'}n_{\dot\gamma},n_{\gamma'})g(n_{\gamma'},e_\epsilon)g(n_{\gamma'},e_\alpha)\Big]_{|(0,0)}\,,\nonumber
\end{eqnarray}
\begin{eqnarray}
\partial_t\partial_s\Lambda_{12}{}^0{}_{\alpha}{}_{|(0,0)}&=&\Big[\frac{g(\nabla_{\dot\gamma}n_{\gamma'},e_\alpha)}{N}g(\nabla_{\gamma'}n_{\dot\gamma},n_{\gamma'})+\frac{g(n_{\gamma'},e_\alpha)}{N}\big(g(\nabla_{\gamma'}n_{\dot\gamma},\nabla_{\dot\gamma}n_{\gamma'})+g(\nabla_{\dot\gamma}\nabla_{\gamma'}n_{\dot\gamma})\big)\nonumber\\
&&\quad + g(\nabla_{\dot\gamma}\nabla_{\gamma'}n_{\dot\gamma},e_\alpha)+\frac{g(n_{\gamma'},e_\alpha)}{N^2}g(\nabla_{\gamma'}n_{\dot\gamma},n_{\gamma'})g(\dot\gamma,\nabla_{\dot\gamma}n_{\gamma'})\Big]_{|(0,0)}\,,\\
\partial_t\partial_s\Lambda_{12}{}^\beta{}_{\alpha}{}_{|(0,0)}&=&-\delta^{\beta\delta}\Big[\frac{2}{N}\Big(g(\nabla_{\dot\gamma}n_{\gamma'},e_{(\alpha})g(e_{\delta)},\nabla_{\gamma'}n_{\dot\gamma})+g(n_{\gamma'},e_{(\alpha})g(e_{\delta)},\nabla_{\dot\gamma}\nabla_{\gamma'}n_{\dot\gamma})\Big)\nonumber\\
&&+\frac{2}{N^2}\Big(g(n_{\gamma'},e_{(\alpha})g(e_{\delta)},\nabla_{\gamma'}n_{\dot\gamma})g(\nabla_{\dot\gamma}n_{\gamma'},\dot\gamma)+g(n_{\gamma'},e_{(\alpha})g(e_{\delta)},\nabla_{\dot\gamma}n_{\gamma'})g(\nabla_{\gamma'}n_{\dot\gamma},n_{\gamma'})\Big)\nonumber\\
&&+\frac{g(n_{\gamma'},e_\alpha)g(n_{\gamma'},e_\delta)}{N^2}\Big(g(\nabla_{\dot\gamma}\nabla_{\gamma'}n_{\dot\gamma},2\dot\gamma+n_{\gamma'})+g(\nabla_{\gamma'}n_{\dot\gamma},n_{\gamma'})g(\nabla_{\dot\gamma}n_{\gamma'},\dot\gamma)\Big.\nonumber\\
&&\qquad\qquad\qquad\qquad\qquad\Big.+g(\nabla_{\gamma'}n_{\dot\gamma},\nabla_{\dot\gamma}n_{\gamma'})\Big)\Big]_{|(0,0)}\,.\nonumber
\end{eqnarray}

Combining these expressions finally yields the components $\Delta\Lambda^\rho{}_\alpha{}_{|(0,0)}$. We obtain
\begin{eqnarray}
\Delta\Lambda^0{}_\alpha{}_{|(0,0)}&=&-\partial_s\partial_t(\Lambda_{12}{}^0{}_{\alpha}(t,s))_{|(0,0)}\Lambda_{11}{}^0{}_{0}(0)-\partial_s\partial_t(\Lambda_{12}{}^\beta{}_{\alpha}(t,s))_{|(0,0)}\Lambda_{11}{}^0{}_{\beta}(0)\nonumber\\
&&-\partial_s(\Lambda_{12}{}^0{}_{\alpha}(t,s))_{|(0,0)}\partial_t(\Lambda_{11}{}^0{}_{0}(t))_{(0,0)}-\partial_s(\Lambda_{12}{}^\beta{}_{\alpha}(t,s))_{|(0,0)}\partial_t(\Lambda-{11}{}^0{}_{\beta}(t))_{(0,0)}\nonumber\\
&=&-g(\nabla_{\dot\gamma}\nabla_{\gamma'}\dot\gamma,e_\alpha)_{|(0,0)}
\end{eqnarray}
as desired. Moreover we calculate
\begin{eqnarray}
\Delta\Lambda^\delta{}_\alpha{}_{(0,0)}&=&-\partial_s\partial_t(\Lambda_{12}{}^0{}_{\alpha}(t,s))_{|(0,0)}\Lambda_{11}{}^\delta{}_{0}(0)-\partial_s\partial_t(\Lambda_{12}{}^\beta{}_{\alpha}(t,s))_{|(0,0)}\Lambda_{11}{}^\delta{}_{\beta}(0)\nonumber\\
&&-\partial_s(\Lambda_{12}{}^0{}_{\alpha}(t,s))_{|(0,0)}\partial_t(\Lambda_{11}{}^\delta{}_{0}(t))_{(0,0)}-\partial_s(\Lambda_{12}{}^\beta_{\alpha}(t,s))_{|(0,0)}\partial_t(\Lambda_{11}{}^\delta{}_{\beta}(t))_{(0,0)}\nonumber\\
&=&\frac{1}{N}(\nabla_{\dot\gamma}\nabla_{\gamma'}\dot\gamma)^\delta g(n_{\gamma'},e_\alpha)+\frac{(\nabla_{\gamma'}\dot\gamma)^\delta}{N^2\sqrt{|g(\gamma',\gamma')|}}g(n_{\gamma'},e_\alpha)g(\nabla_{\dot\gamma}\gamma',n_{\gamma'}+\dot\gamma)\\
&&-\frac{n_{\gamma'}^\delta}{N}\Big(g(\nabla_{\dot\gamma}\nabla_{\gamma'}\dot\gamma,e_\alpha)+\frac{g(\nabla_{\dot\gamma}\gamma',e_\alpha)}{N\sqrt{|g(\gamma',\gamma')|}}g(\nabla_{\dot\gamma}\gamma',n_{\gamma'}+\dot\gamma)\Big)\,.\nonumber
\end{eqnarray}
Observing that $\nabla_{\gamma'}\dot\gamma = \nabla_{\dot\gamma}\gamma'$ due to vanishing torsion, and inserting the abbreviations $v=\nabla_{\dot\gamma}\gamma'$ and $a=\nabla_{\dot\gamma}\nabla_{\dot\gamma}\gamma'$, this is precisely the result stated in equation~(\ref{eq:dL}). This finally concludes the proof of Theorem~1. $\square$

\section{Kerr spacetime curvature functions}
Here we display the explicit form of the functions from which the components of the Riemann curvature tensor in Kerr spacetime, which are relevant for the experiment described in section \ref{sec:kerr}, are built:
\begin{eqnarray}\label{app:kerr}
C_{0}&=&9 a^4 + 8 a^2 r^2 - 8 r^4 + 4 a^2 (3 a^2 + 2 r^2) \cos(2 \theta) + 3 a^4 \cos(4 \theta),\\
C_{1}&=&2(\Sigma+2a^2\sin^2\theta),\\
C_2&=&-2 a (a^2 + r^2) \sin^2\theta,\\
D_0&=&3 a^4 - 8 a^2 r^2 - 24 r^4 + 4 a^2 (a^2 - 2 r^2) \cos(2 \theta) + a^4 \cos(4 \theta),\\
D_1&=&10 a^2 + 8 r (-r_s + r) - 2 a^2 \cos(2 \theta),\\
D_2&=&-8 a (3 a^2 + r (-2 r_s + 3 r)) \sin^2\theta,\\
D_3&=&-4 (-2 a^4 + a^2 (r_s - 3 r) r - r^4 +  a^2 (a^2 + r (-r_s + r)) \cos(2 \theta)) \sin^2(\theta),
\end{eqnarray}

\end{document}